\documentclass[aps,prb,twocolumn,superscriptaddress,floatfix,longbibliography,citeautoscript]{revtex4-2}

\usepackage{amsmath,amssymb} 
\usepackage{bm} 
\usepackage{graphicx}
\usepackage{comment} 
\usepackage[normalem]{ulem} 

\usepackage{minted}
\usepackage{textcomp} 

\usepackage{tikz}
\usetikzlibrary{quantikz2}

\usepackage{enumitem}
\setlist{noitemsep,leftmargin=*,topsep=0pt,parsep=0pt}

\usepackage{xcolor}
\definecolor{lightgray}{gray}{0.6}
\definecolor{medgray}{gray}{0.4}
\definecolor{mRed}{RGB}{230, 0, 50}
\colorlet{newtextColor}{mRed}

\usepackage[hidelinks]{hyperref} 
\newif\ifptitle
\newif\ifpnumber
\newcounter{para}

\newif\iftrackchanges

\usepackage{mdframed}
\newmdenv[
  linecolor={\iftrackchanges newtextColor\else white\fi},
  linewidth=2pt,
  topline=false,
  bottomline=false,
  rightline=false,
  skipabove=\topsep,
  skipbelow=\topsep,
  leftmargin=-12pt,
  innertopmargin=0pt,
  innerbottommargin=0pt
]{newtextblock}

\usepackage{float}

\newcommand{\affrqc}{Russian Quantum Center, Moscow 143025, Russia}
\newcommand{\afflebedev}{P.N. Lebedev Physical Institute of the Russian Academy of Sciences, Moscow 119991, Russia}
\newcommand{\affmipt}{Moscow Institute of Physics and Technology, Dolgoprudny 141700, Russia}
\newcommand{\affmisis}{National University of Science and Technology ``MISIS'', Moscow 119049, Russia}
\newcommand{\affsteklov}{Steklov Mathematical Institute of Russian Academy of Sciences, Moscow 119991, Russia}

\begin{document}

\title{Experimental characterization of the Toffoli gate via channel spectrum benchmarking}

\author{D.~K. Korliakov}
\email[]{korliakov.dk@phystech.edu}
\affiliation{\affrqc}
\affiliation{\affmipt}

\author{B.~I. Bantysh}
\affiliation{\affrqc}

\author{A.~S. Borisenko}
\affiliation{\afflebedev}

\author{I.~V. Zalivako}
\affiliation{\afflebedev}

\author{E.~O. Kiktenko}
\affiliation{\affrqc}
\affiliation{\affsteklov}
\affiliation{\affmisis}

\begin{abstract}
Channel spectrum benchmarking (CSB) provides a robust framework for characterizing quantum gate fidelities while remaining insensitive to state preparation and measurement (SPAM) errors. Yet, current CSB implementations encounter fundamental challenges when reconstructing noisy eigenvalues, particularly in the presence of spectral degeneracies and off-diagonal noise components in the target gate’s eigenbasis. These issues become especially pronounced in the strong noise regime for gates with fidelities around $90\%$. To address these limitations, we introduce an extended CSB model together with a fidelity estimate interval (FEI) -- an interval-valued estimate of the target gate fidelity. Numerical simulation demonstrates that FEI remains sufficiently narrow, with its midpoint reliably approximating the true fidelity. We further validate the protocol on a trapped-ion quantum processor by benchmarking two implementations of the three-qubit Toffoli gate. The results reveal a clear advantage of the qutrit-based implementation over its qubit-based counterpart.
\end{abstract}

\maketitle

\section{Introduction}

Performance limitations in today's quantum computers stem largely from quantum noise~\cite{preskill2018quantum}, which introduces errors during computations and reduces the reliability of quantum operations. This noise arises from various sources, including qubits environment, imperfect control pulses, and unwanted interactions between qubits. Since quantum systems are highly sensitive, even minor disturbances can significantly degrade computational accuracy. 

A comprehensive understanding of the noise structure in quantum processors is essential for the development of scalable and reliable quantum computation. To achieve this, a diverse set of characterization techniques has been developed to precisely quantify and benchmark the performance of fundamental quantum operations, including state initialization~\cite{lin2021independent, sotirova2024high, khan2025separate}, single- and two-qubit gate implementations~\cite{li2023error, smith2025single, chen2025efficient, wang2020high, fang2022crosstalk, shapira2023robust}, measurement procedures~\cite{lin2021independent, sotirova2024high, khan2025separate}, and qubit reset protocols~\cite{egger2018pulsed, basilewitsch2019reservoir, mi2022securing}. These methods enable rigorous validation of operational fidelities and identification of error sources critical for error mitigation strategies. Current characterization techniques, such as randomized benchmarking (RB)~\cite{knill2008randomized, wallman2014randomized, magesan2011scalable, proctor2019direct}, interleaved RB~\cite{magesan2012efficient}, mirror RB~\cite{proctor2022scalable}, cycle benchmarking (CB)~\cite{erhard2019characterizing}, quantum process tomography (QPT)~\cite{chuang1997prescription, merkel2013self, altepeter2003ancilla, mohseni2008quantum, o2004quantum, white2022non}, gate set tomography (GST)~\cite{blume2013robust, nielsen2021gate, greenbaum2015introduction, blume2017demonstration}, either focus on specific gate subsets, cannot directly characterize individual gate performance or have significant drawbacks such as sensitivity to state preparation and measurement (SPAM) errors.

The channel spectrum benchmarking (CSB)~\cite{gu2023benchmarking} is a promising technique for fidelity estimation of experimental implementations of individual quantum gates and quantum circuits. Unlike randomized benchmarking, which provides an average fidelity over the gates of the Clifford group, or cyclic benchmarking, restricted to gates $U$ with property $U^m=I$, CSB imposes almost no constraints on target gate. Additionally, CSB is robust against SPAM errors, which often distort fidelity estimates.

This work addresses a critical challenge in CSB: the inability to reconstruct the complete set of noisy eigenvalues of the target operation for fidelity estimation. We demonstrate that CSB protocol, proposed in~\cite{gu2023benchmarking}, struggles due to high degeneracy in the target operation’s eigenvalue spectrum and off-diagonal noise elements in the target operation’s eigenbasis. These issues create ambiguities in eigenvalue reconstruction, leading to inaccurate single point fidelity estimates. To resolve this, we propose an alternative CSB method incorporating more reliable processing methods, including models that account for off-diagonal noise elements and subsequent filtering of the obtained eigenvalue estimates. As an alternative to a single point estimate, our CSB protocol yields a fidelity estimate interval (FEI) -- an interval estimate of fidelity. Unlike a single point estimate, FEI mitigates the impact of ambiguities in reconstructing the target operation's eigenvalues. We demonstrate that in CSB protocol simulations for fidelity estimation of a three-qubit Toffoli gate, the obtained FEI is sufficiently narrow, and its midpoint -- which can be considered a single point estimate -- closely approximates the true fidelity value.

We experimentally evaluate the fidelity of a three-qubit Toffoli gate on an ion-based quantum platform, comparing qubit and qutrit~\cite{nikolaeva2022decomposing, nikolaeva2025scalable} implementations. The results show comparable FEIs for both implementations.

This work is organized as follows. In Sec.~\ref{sec:Preliminaries}, we discuss the mathematical concepts needed to describe the protocol. In Sec.~\ref{sec:CSB_protocol}, we detail the CSB protocol, including its theoretical foundations and practical implementation steps. In Sec.~\ref{sec:Eigenvalues extraction}, we introduces modifications of the standard CSB. In Sec.~\ref{sec:Simulation}, we present simulation results validating the CSB method. In Sec.~\ref{sec:Experiment}, we discuss experimental results comparing qubit and qutrit implementations of the Toffoli gate. Finally, in Sec.~\ref{sec:Conclusion_and_discussion}, we summarize our findings.

\begin{figure*}[!htb]
    \centering
    \includegraphics[width=0.98\linewidth]{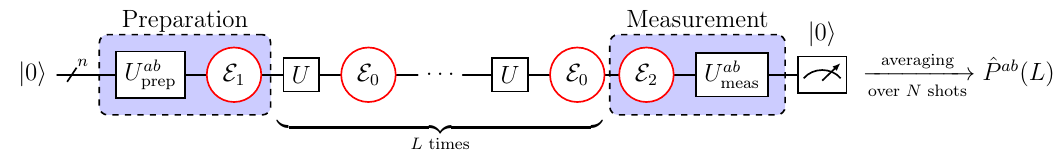}
    \caption{Schematic quantum circuit for the CSB protocol. Quantum gates are represented by black rectangles. Red circles indicate the circuit locations where we model the presence of inherent quantum noise.}
    \label{fig:CSB_scheme}
\end{figure*}

\section{Preliminaries}\label{sec:Preliminaries}

This section introduces the necessary concepts of quantum channels, their representations, and properties to describe the protocol. Consider a quantum gate $U$ acting on a $d$-dimensional Hilbert space, represented by a $d \times d$ unitary matrix. In experimental settings, quantum systems are susceptible to noise, making the practical implementation non-unitary. Such implementations are described using quantum channels, completely-positive and trace-preserving (CPTP) maps~\cite{nielsen2010quantum}.

The noisy implementation of gate $U$ corresponds to a quantum channel $\mathcal{\tilde{U}} = \mathcal{E} \circ \mathcal{U}$, where $\mathcal{U}$ represents the ideal gate operation and $\mathcal{E}$ denotes the noise channel. We restrict our analysis to Markovian noise, where errors are uncorrelated with each other and can be associated with individual gate operations. Quantum channels can be represented in several equivalent ways. For instance, the action of any quantum channel $\mathcal{R}$ can be expressed using Kraus operators $\{R_i\}$~\cite{kraus1971general} as:
\begin{equation}
    \mathcal{R}(\rho) = \sum_i R_i \rho R_i^{\dagger},
\end{equation}
where $\rho$ is an arbitrary density operator. For a unitary operation $U$, the minimal Kraus representation consists of a single operator $\{U\}$.

We employ the matrix representation for quantum channels, where each channel $\mathcal{R}$ corresponds to a $d^2 \times d^2$ matrix $G_{\mathcal{R}}$ acting on a vectorized density operator $|\rho \rangle \rangle$:
\begin{equation}
    |\rho \rangle \rangle \xrightarrow{\mathcal{R}} G_{\mathcal{R}} |\rho \rangle \rangle.
\end{equation}
Here, $|\rho \rangle \rangle$ denotes the vector formed by stacking the columns of $\rho$ and $G_{\mathcal{R}} = \sum_i R_i^{*} \otimes R_i$. The ideal unitary channel $\mathcal{U}$ corresponds to the matrix $G_{\mathcal{U}} = U^{*} \otimes U$.

Using the spectral decomposition of the unitary operator $U = A \Lambda A^{\dagger}$, the matrix $G_{\mathcal{U}}$ can be constructed as follows:
\begin{equation}\label{eq:spectral_dec_G}
    G_{\mathcal{U}} = \underbrace{(A^{*} \otimes A)}_{C} \cdot \underbrace{(\Lambda^* \otimes \Lambda)}_{K_{\mathcal{U}}} \cdot \underbrace{(A^{T} \otimes A^{\dagger})}_{C^{\dagger}}.
\end{equation}
Consequently, the matrix representation of any quantum channel $\mathcal{R}$ can be expressed in the eigenbasis of $G_{\mathcal{U}}$ as $G_{\mathcal{R}} = C K_{\mathcal{R}} C^{\dagger}$.

The similarity between the ideal channel $\mathcal{U}$ and its noisy implementation $\tilde{\mathcal{U}}$ is quantified by the process fidelity~\cite{kliesch2021theory}:
\begin{equation}\label{eq:fidelity}
    \mathcal{F}\left( \mathcal{U}, \tilde{\mathcal{U}} \right) = \frac{\text{tr} \left( G_{\mathcal{U}}^{\dagger} G_{\tilde{\mathcal{U}}} \right)}{d^2} = \frac{\text{tr} \left( K_{\mathcal{E}} \right)}{d^2} = \frac{\sum_i \lambda_{\mathcal{E}}^i}{d^2},
\end{equation}
where $K_{\mathcal{E}} = C^{\dagger} G_{\mathcal{E}} C$ is the matrix representation of the noise channel in the eigenbasis of $G_{\mathcal{U}}$. 

The estimate of the fidelity requires knowledge of the average diagonal elements of the noise channel matrix:
\begin{equation}\label{eq:fidelity_est}
    \hat{\mathcal{F}}(\mathcal{U}, \tilde{\mathcal{U}}) = \langle \lambda_{\mathcal{E}}^i \rangle = \langle \mathrm{Re}(\lambda_{\mathcal{E}}^i) \rangle,
\end{equation}
where the second equality follows from the complex-conjugate structure of the diagonal elements of the noise matrix  $K_{\mathcal{E}}$.

\section{CSB protocol}\label{sec:CSB_protocol}

This section presents a concise description of the channel spectrum benchmarking (CSB) protocol, following the original formulation~\cite{gu2023benchmarking}.

Consider the circuit shown in Fig.~\ref{fig:CSB_scheme}. The circuit can be divided into three logical blocks: 

\begin{enumerate}
    \item Preparation of the state 
    \begin{equation*}
        |\psi_{ab}\rangle = \begin{cases}
            \frac{|\psi_a\rangle + |\psi_b\rangle}{\sqrt{2}}, \; & \text{if $a \neq b$}\\
            |\psi_{a}\rangle, &\text{if $a = b$}
        \end{cases},
    \end{equation*}
    where $|\psi_{a}\rangle$ and $|\psi_{b}\rangle$ are eigenstates of the target gate $U$;
    \item Application of the target gate $U$ repeated $L$ times;
    \item Measurement in the eigenbasis of $U$ (the state $|\psi_{ab} \rangle$ is of interest).
\end{enumerate}

In the circuit, gate $U^{ab}_{\text{prep}}$ prepares state $|\psi_{ab} \rangle$, while $U^{ab}_{\text{meas}} = \left(U^{ab}_{\text{prep}}\right)^{\dagger}$. Thus, the probability $P^{ab}(L)$ of measuring state $|\psi_{ab} \rangle$ after applying the target operation takes the form:
\begin{multline}\label{eq:P(L)_huge_form}
    P^{ab}(L) = \langle \langle 0| G^{ab}_{\text{meas}}  G_{\mathcal{E}_2}   \left( G_{\mathcal{E}_0} G_{\mathcal{U} } \right)^L G_{\mathcal{E}_1} G^{ab}_{\text{prep}} |0 \rangle \rangle  = \\
    =\langle \langle \rho_{ab}| C \left( K_{\mathcal{E}_2}   \left( K_{\mathcal{E}_0} K_{\mathcal{U} } \right)^L K_{\mathcal{E}_1} \right) C^{\dagger} |\rho_{ab} \rangle \rangle = \\
    =\langle \langle ab|  K_{\mathcal{E}_2}   \left( K_{\mathcal{E}_0} K_{\mathcal{U} } \right)^L K_{\mathcal{E}_1} |ab \rangle \rangle.
\end{multline}
Here, $K_{\mathcal{U}}$ (defined in Eq.~\eqref{eq:spectral_dec_G}) is diagonal with elements $\lambda^i_{\mathcal{U}}$. The vector $|ab \rangle \rangle$ has dimension $d^2$ with non-zero components exclusively at indices $\{aa, ab, ba, bb\}$. Under the weak noise approximation, which assumes negligible off-diagonal elements in $K_{\mathcal{E}_k}$, Eq.~\eqref{eq:P(L)_huge_form} simplifies to:
\begin{equation}\label{eq:P(L)_small_form}
    P^{ab}(L) 
    \approx \sum_{i \in \{ aa, ab, ba,bb  \}} r_i \left( \lambda_{\tilde{\mathcal{U}}}^{i}\right)^L
\end{equation}
with coefficients $r_i$ that appear in complex-conjugate pairs. Fitting the experimental data with this equations allows extracting noisy eigenvalues $\lambda_{\tilde{\mathcal{U}}}^{i}$. In the original proposal authors employ matrix pencil (MP) method~\cite{gu2023benchmarking}. In this work, we rely on numerical optimization, since it was found more stable for different gates and noise cases.

After that, one can finally extract the eigenvalues of the noisy channel using the linear approximation:
\begin{equation}\label{eq:linear_lambda}
    \lambda_{\tilde{\mathcal{U}}}^{i} = \lambda_{\mathcal{E}}^{i}\lambda_{\mathcal{U}}^{i}
\end{equation}

Thus, the CSB protocol proceeds as follows:
\begin{enumerate}
    \item Obtaining estimates of dependencies $\hat{P}^{ab}(L)$:
    \begin{enumerate}
        \item Preparation of the state $|\psi_{ab}\rangle$;
        \item Sequential application of the gate $U$ to the prepared state $L$ times within the range ${L\in[0, L_{\max}]}$;
        \item Measurement in the eigenbasis of gate $U$;
        \item Repeating the previous steps $N$ times to accumulate statistics and average measurement results for fixed $L$ and initial state to obtain $\hat{P}^{ab}(L)$;
        \item Varying $L$ within the specified range and repeating all previous steps;
        \item Preparing a different initial state and repeating all previous steps.
    \end{enumerate}
    \item Processing experimental data:
    \begin{enumerate}
        \item Fitting experimental dependencies $\hat{P}(L)$ with a chosen model and extracting estimates of channel eigenvalues $\lambda_{\tilde{\mathcal{U}}}^{i}$;
        \item Computing noise eigenvalues in linear approximation $\lambda_{\mathcal{E}}^{i} = \lambda_{\tilde{\mathcal{U}}}^{i}/\lambda_{\mathcal{U}}^{i}$.
    \end{enumerate}
\end{enumerate}
The process fidelity is then estimated from the obtained noise eigenvalues using the following equation:
\begin{equation}\label{eq:fid_degenerate}
    \hat{\mathcal{F}} = \frac{d_{\text{ts}} \langle \lambda_{\mathcal{E}}^{ab} \rangle |_{\lambda_{\mathcal{U}}^{ab} = 1}     + d_{\text{ns}} \langle \lambda_{\mathcal{E}}^{ab} \rangle |_{\lambda_{\mathcal{U}}^{ab} \neq 1}   }{d^2},
\end{equation}
where $d_{\text{ts}}$ denotes the multiplicity of eigenvalue $\lambda_{\mathcal{U}} = 1$, and $d_{\text{ns}} = d^2 - d_{\text{ts}}$. This weighted averaging formula accounts for the non-uniform distribution of eigenoperator subspaces in the channel representation. Since the trivial subspace ($\lambda_{\mathcal{U}}^{ab} = 1$) and non-trivial subspace ($\lambda_{\mathcal{U}}^{ab} \neq 1$) typically have different dimensions, the formula ensures proper weighting of their contributions according to their respective multiplicities in the full operator space. This formula can be used with both uniform sampling of initial states $|\psi_{ab}\rangle$, as proposed in~\cite{gu2023benchmarking}, and with their complete enumeration, as implemented in this work.

\section{Strong noise CSB}\label{sec:Eigenvalues extraction}

In this work we focus on strong noise regime (fidelity about 90\%) and address multiple effects that may lower the fidelity estimation precision.

First, in this case the 4-term model~\eqref{eq:P(L)_small_form} becomes unreliable, since off-diagonal terms of the matrix $K_{\mathcal{E}_k}$ alter the resulting dependency. Instead, we consider a slightly more complex 6-term model
\begin{equation}\label{eq:g_model}
    P^{ab}(L) = \sum_{i=0}^{5} f_i \left(z^i\right)^L,
\end{equation}
ensuring that all $z^i$ appear in complex conjugate pairs. However, this may introduce some additional eigenvalues that should be filtered out.
\begin{enumerate}
    \item Values $z^i$ with small $|f_i|$ are removed;
    \item Values $z^i$ whose complex phases deviate from the complex phases of the corresponding noiseless eigenvalues $\lambda_{\mathcal{U}}^j$ by more than a threshold $\delta$ are removed. The threshold value $\delta$ depends on the target operation and is determined empirically.
\end{enumerate}
In the Sec.~\ref{sec:Simulation} we will demonstrate that this simple improvement leads to a better fit of numerical experimental data and fidelity estimation precision.

Next, the degeneracy of the spectrum of the investigated operation may also complicate the recovery of diagonal elements. Under the influence of noise, the same degenerate eigenvalue $\lambda_{\mathcal{U}}^i$ may split into several noisy eigenvalues. Moreover, when processing experimental data, it remains challenging to determine whether those eigenvalues are actually split or whether their estimates differ due to statistical averaging noise.

Rather than addressing these challenges directly, our approach circumvents them by introducing the fidelity estimate interval (FEI) as an alternative metric. The FEI construction involves generating multiple fidelity estimates from subsamples of size $d^2$ drawn from $\{\mathrm{Re}(\lambda_{\mathcal{E}}^i)\}$ with replacement. Sampling with replacement accommodates uncertainties in the frequency of specific $\lambda_{\mathcal{E}}^i$ estimates. The real part is used to account for complex-conjugate pairs in the dataset. The FEI boundaries are determined as quantiles of the fidelity distribution across subsamples. The single point estimate derived from the full set $\{z^i\}$ corresponds to the FEI midpoint.

The algorithm for finding the FEI is shown below:

\begin{enumerate}
    \item The set $\{z^i\}$ is converted to a set of estimates $\{\mathrm{Re}(\lambda_{\mathcal{E}}^i)\})$ according to Eq.~\eqref{eq:linear_lambda};
    \item From the set $\{\mathrm{Re}(\lambda_{\mathcal{E}}^i)\}$, $N$ samples of size $d^2$ are randomly drawn with replacement;
    \item For each sample, a fidelity estimate $\{\hat{F}^*_{(k)}\}_{k=1}^N$ is computed using Eq.~\eqref{eq:fidelity_est};
    \item The distribution of fidelity estimates $\hat{\mathcal{F}}^*$ is constructed from the obtained $N$ estimates;
    \item The FEI is calculated as:

\begin{equation}
    \text{FEI} = [\hat{\mathcal{F}}^*_{0.025}; \hat{\mathcal{F}}^*_{0.975}],
\end{equation}
where $\hat{\mathcal{F}}^*_{p}$ is the $p$-th quantile of the fidelity distribution;

\end{enumerate}

\section{Simulation}\label{sec:Simulation}

This section describes the Toffoli gate and its properties relevant for CSB experiments, and presents computer simulation results of the CSB procedure.

The three-qubit Toffoli gate exhibits high spectral degeneracy. It possesses eigenvalue $\Lambda_{-} = -1$ with multiplicity 1 (corresponding to eigenstate $|11-\rangle$) and eigenvalue $\Lambda_{+} = +1$ with multiplicity 7 (corresponding to eigenstates $\{|000\rangle, |001\rangle, |010\rangle, |011\rangle, |100\rangle, |101\rangle, |11+\rangle\}$).

Following Eq.~\eqref{eq:spectral_dec_G}, the Toffoli channel's matrix representation consequently exhibits two eigenvalues: $\lambda_{-} = -1$ with multiplicity 14 and $\lambda_{+} = +1$ with multiplicity 50. As discussed in Sec.~\ref{sec:Eigenvalues extraction}, this spectral degeneracy introduces reconstruction ambiguities for noisy eigenvalues.

In simulations, the three-qubit Toffoli gate served as the target operation, with noise modeling incorporating single-qubit amplitude damping and small unitary rotations around the Bloch sphere's Y-axis. The target operation was implemented as a single three-qubit gate, with noise affecting both the gate and state preparation/measurement operations. Probability estimates were obtained through 1000-shot averaging, with maximum repetition depth $L_{\max} = 40$.

We compared our processing method (Sec.~\ref{sec:Eigenvalues extraction}) with the approach proposed in~\cite{gu2023benchmarking}. This approach fits experimental data using the matrix pencil method with a 4-term model~\eqref{eq:P(L)_small_form}, without subsequent filtering of the obtained estimates $\{\lambda_{\tilde{\mathcal{U}}}^i\}$. A single point fidelity estimate is obtained using a Eq.~\eqref{eq:fid_degenerate}. The reference implementation of this approach is available as open-source code at GitHub repository~\cite{yanwu_gu_2023_8310716}. 

\begin{figure}[!htb]
    \centering
    \includegraphics[width=0.98\linewidth]{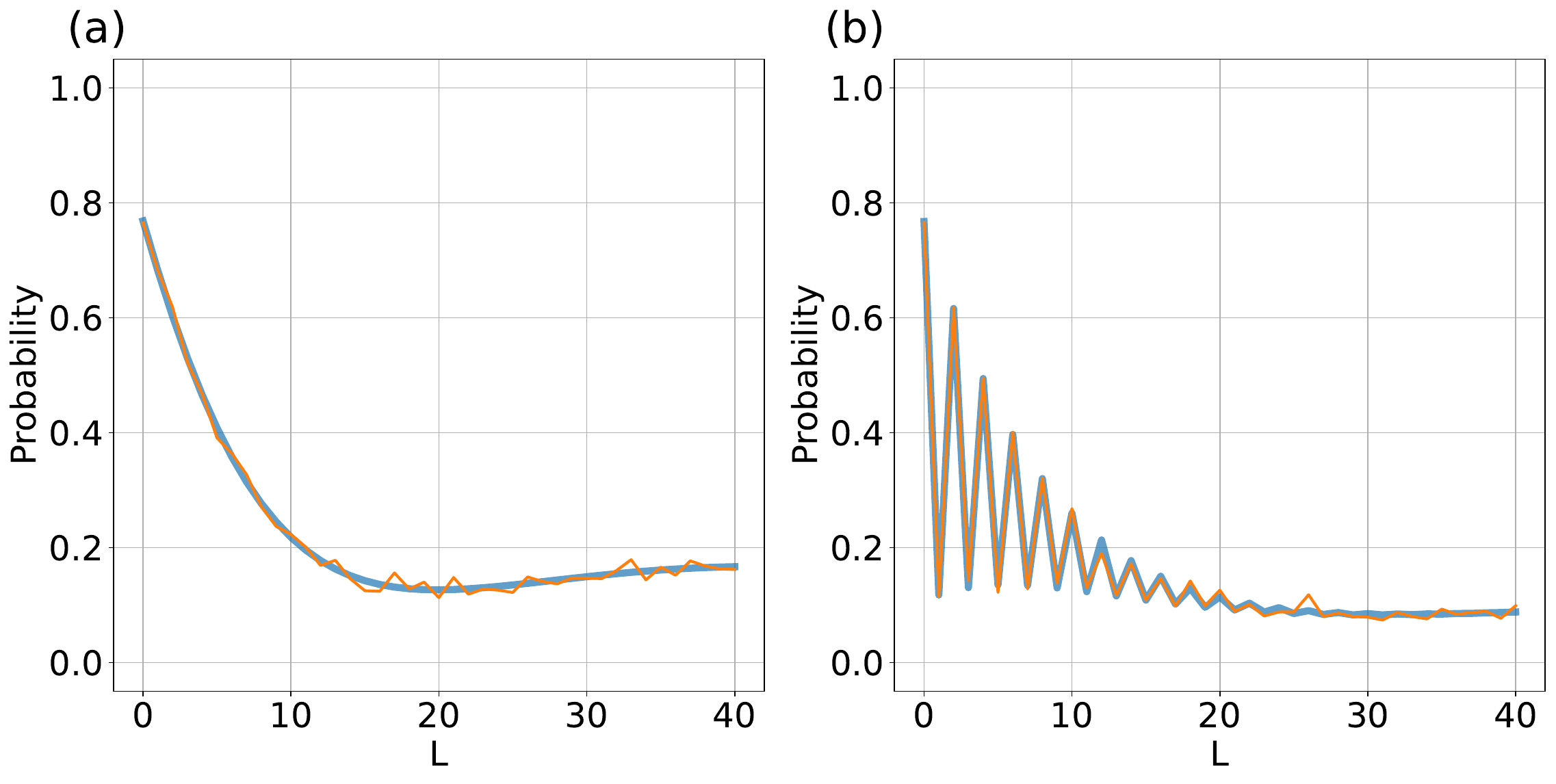}
    \caption{Typical $\hat{P}^{ab}(L)$ dependencies from CSB simulations. (a) State $(|101\rangle + |010\rangle)/\sqrt{2}$ shows smooth decay; (b) State $(|100\rangle + |11-\rangle)/\sqrt{2}$ exhibits oscillations. Orange lines: simulated data; blue curves: fitted 6-term models~\ref{eq:g_model}.}
    \label{fig:oscill}
\end{figure}

Simulations produced numerical experimental data $\hat{P}^{ab}(L)$ showing two distinct patterns (Fig.~\ref{fig:oscill}): smooth decay (a) and oscillatory behavior (b). 

Oscillations originate from terms of form $(r_i ((-1)\cdot \lambda_{\mathcal{E}}^i)^L + \text{h.c.})$ in Eq.~\eqref{eq:g_model}, arising when prepared states $|\psi_{ab}\rangle = (|\psi_a\rangle + |\psi_b\rangle)/\sqrt{2}$ combine eigenstates with different eigenvalues ($\lambda_a \neq \lambda_b$).

Fig.~\ref{fig:oscill}(b) displays the estimated dependency $\hat{P}^{ab}(L)$ for state $|\psi\rangle = (|100\rangle + |11-\rangle)/\sqrt{2}$, where $|100\rangle$ corresponds to eigenvalue $\Lambda_{+}$ and $|11-\rangle$ to $\Lambda_{-}$. The oscillatory component in Eq.~\eqref{eq:g_model} originates from eigenoperators $|100\rangle \langle 11- |$ and $|11-\rangle \langle100|$ in $G_{\mathcal{U}}$, associated with eigenvalue $\lambda_{-} = -1$. No oscillations appear in Fig.~\ref{fig:oscill}(a) since states $|101\rangle$ and $|010\rangle$ possess identical eigenvalues.

For arbitrary gates, states $|\psi_{ab}\rangle = (|\psi_a\rangle + |\psi_b\rangle)/\sqrt{2}$ produce oscillatory contributions of form $(r_{ab} (\lambda_{\mathcal{E}}^{ab} (\lambda_a \lambda_b^*))^L + \text{h.c.})$, with oscillation frequency $|\text{arg}(\lambda_a) - \text{arg}(\lambda_b)|$.

\begin{figure}[!htb]
    \centering
    \includegraphics[width=0.98\linewidth]{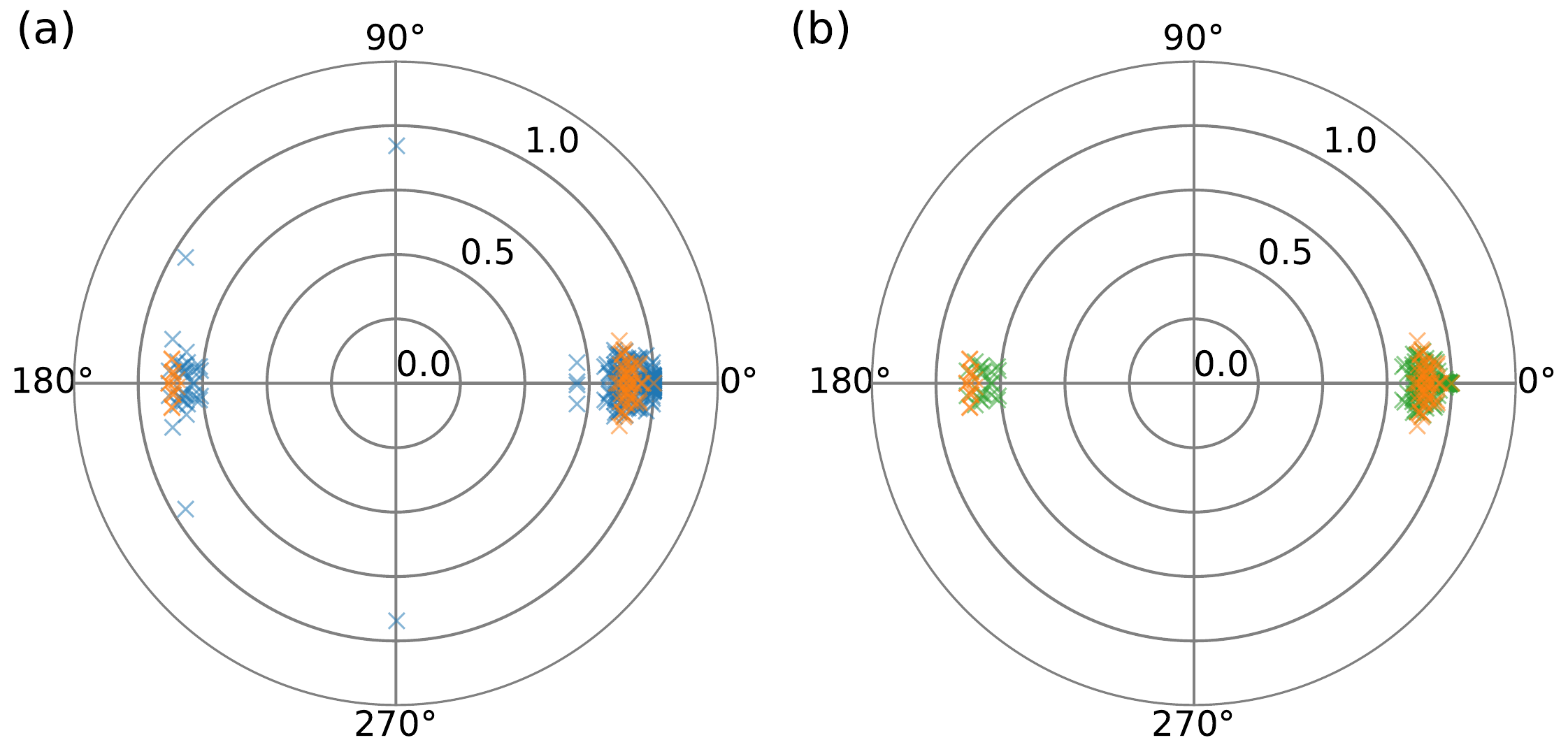}
    \caption{Noisy eigenvalues of the Toffoli gate in the complex plane. Orange: theoretical values; blue: reconstructed estimates obtained by fitting the 6-term model~\eqref{eq:g_model}; green: estimates after applying the filtering procedure proposed in this work.}
    \label{fig:lambds_polar}
\end{figure}

Numerical experimental data $\hat{P}^{ab}(L)$ were fitted using model~\eqref{eq:g_model}, yielding parameter set $\{z^i\}$. After filtering, we obtained estimates $\{\lambda_{\tilde{\mathcal{U}}}^{i}\}$. Fig.~\ref{fig:lambds_polar} compares these estimates with theoretical values, demonstrating significant improvement after filtering.

\begin{figure}[!htb]
    \centering
    \includegraphics[width=0.98\linewidth]{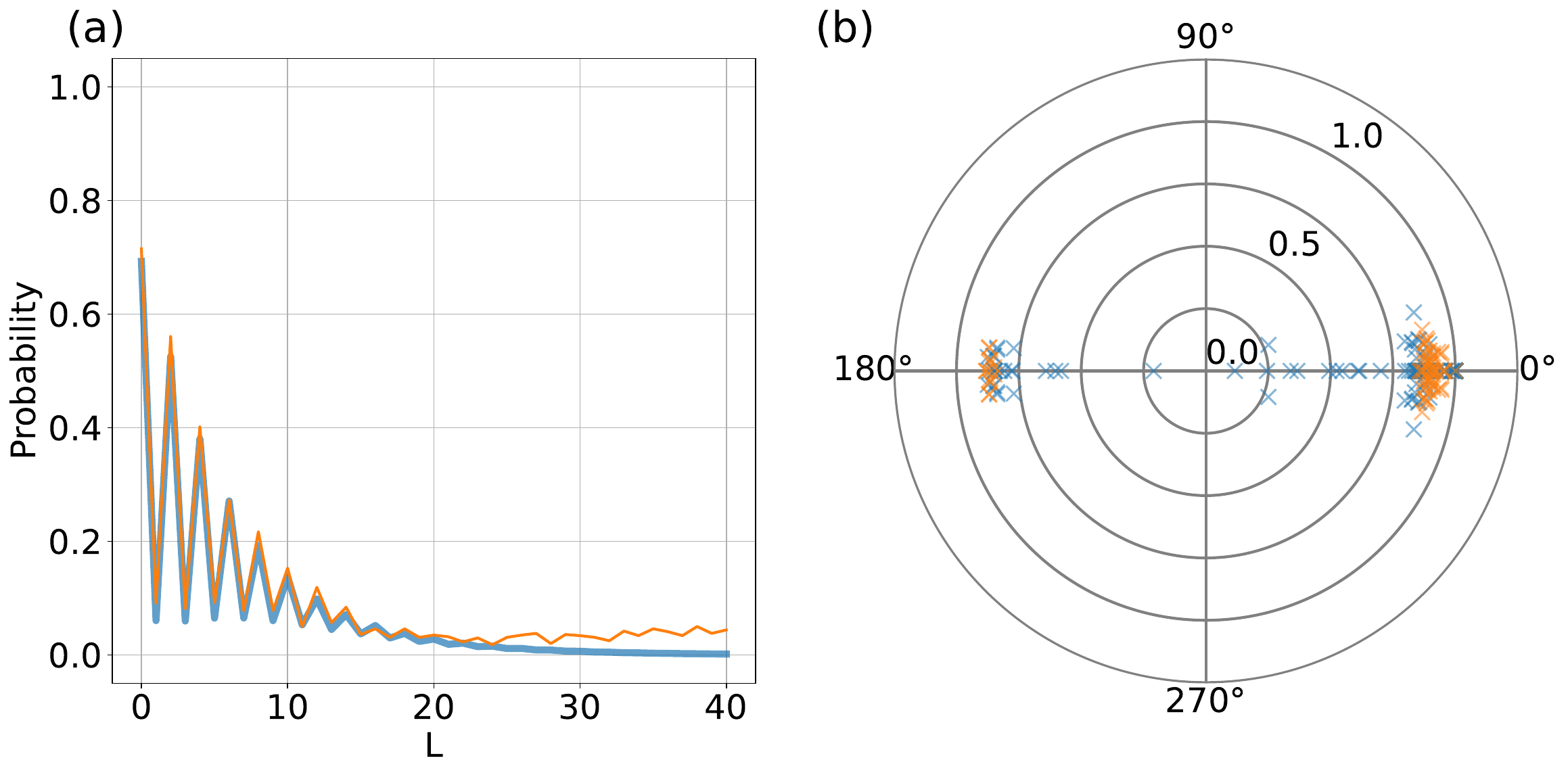}
    \caption{Results of the algorithm proposed in~\cite{gu2023benchmarking}. (a) The dependency $\hat{P}^{ab}(L)$ for state $(|101\rangle + |11-\rangle)/\sqrt{2}$ (orange line: simulated data; blue curve: 4-term model~\eqref{eq:P(L)_small_form} fitted with MP method). (b) Reconstructed noisy eigenvalues of the Toffoli gate in the complex plane (orange: theoretical values; blue: raw estimates obtained using the approach from~\cite{gu2023benchmarking}).}
    \label{fig:MP_lambdas_and_plots}
\end{figure}

Fig.~\ref{fig:MP_lambdas_and_plots} presents the results of the algorithm proposed in~\cite{gu2023benchmarking}. The 4-term model~\eqref{eq:P(L)_small_form} fitted using the MP method demonstrates suboptimal data approximation. Divergences appear at large $L$ values. Furthermore, the reconstruction of noisy eigenvalues $\lambda_{\tilde{\mathcal{U}}}$ is inaccurate, with numerous estimates deviating significantly from theoretical values. This behavior arises because the 4-term model~\eqref{eq:P(L)_small_form} neglects off-diagonal noise elements, which substantially impact fitting quality.

\begin{figure}[!htb]
    \centering
    \includegraphics[width=0.98\linewidth]{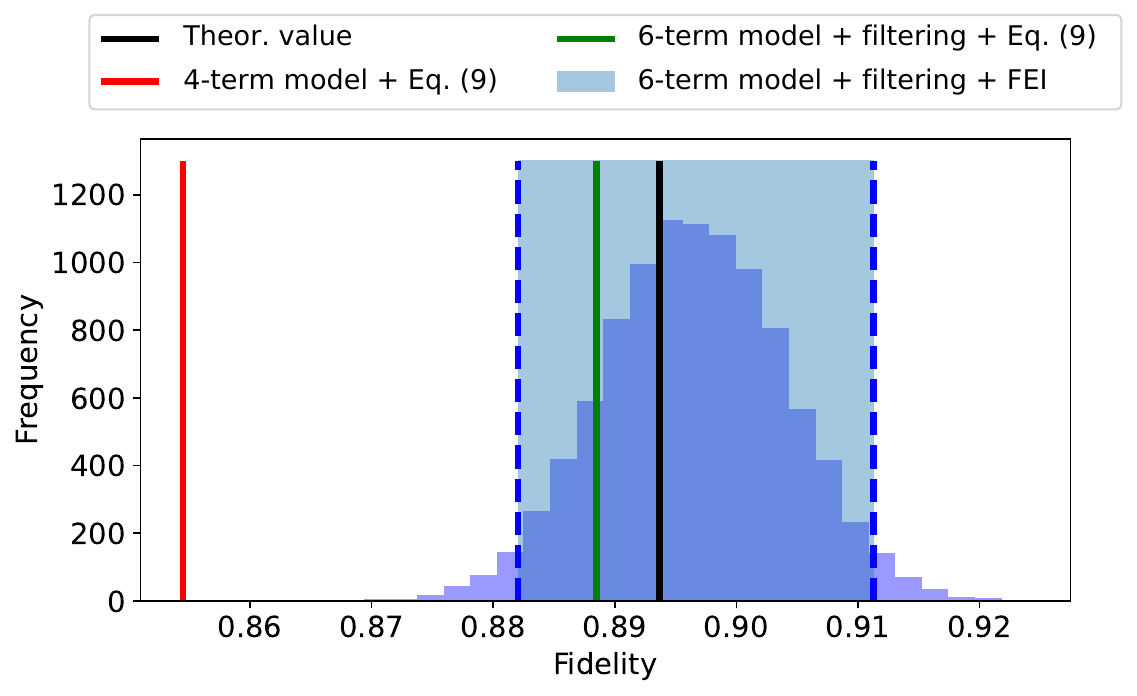}
    \caption{Comparison of fidelity estimation methods for the Toffoli gate. The distribution shows fidelity estimates obtained using: our approach with 6-term model, numerical optimization, filtering, and FEI construction (blue); the same method but using Eq.~\eqref{eq:fid_degenerate} for point estimation (green); the theoretical value (black); and the MP method with 4-term model and Eq.~\eqref{eq:fid_degenerate} (red).}
    \label{fig:MP_this_comp}
\end{figure}

Simulation results yielded $\text{FEI}_{\text{this work}}= [0.882;0.911]$ (single point fidelity estimate $\hat{\mathcal{F}}_{\text{this work}} = 0.897$), matching theoretical prediction $\hat{\mathcal{F}}_{\text{theor}} = 0.893$. The matrix pencil method produced $\hat{\mathcal{F}}_{\text{MP}} = 0.825$. We also calculated a single-point estimate $\hat{\mathcal{F}}_{\text{degenerate}} = 0.888$ using Eq.~\eqref{eq:fid_degenerate} with the 6-term model fitted via numerical optimization and subsequent filtering. Although this estimate shows a minor deviation from the FEI midpoint $\hat{\mathcal{F}}_{\text{this work}}$, it remains within the fidelity estimate interval. Thus, Eq.~\eqref{eq:fid_degenerate} proves particularly useful when employing sampling of initial states rather than complete enumeration. Fig.~\ref{fig:MP_this_comp} compares these results.

\section{Experiment}\label{sec:Experiment}

This section presents experimental CSB results for fidelity estimation of the Toffoli gate on a qudit-based trapped-ion quantum processor. The details on the experimental setup can be found in~\cite{zalivako2025towards, kolachevsky2025quantum}. The processor is based on a chain of $^{171}\textrm{Yb}^+$ ions confined in a linear Paul trap. In the experiments described here the quantum register consisted of ten ions, each encoding a qutrit in states 
\begin{align*}
|0\rangle&=|{}^2S_{1/2}(F=0,m_F=0)\rangle, \\
|1\rangle&=|{}^2D_{3/2}(F=2,m_F=0)\rangle, \\
|2\rangle&=|{}^2S_{1/2}(F=1,m_F=0)\rangle.
\end{align*}
The native gate set of the processor includes single-qudit operations 
\begin{align*}
R_{\phi}^{0j}(\theta)&=\exp(-i\sigma^{0j}_{\phi}\theta/2), \\
R_{z}^{j}(\theta)&=\exp\left(i\theta\ket{j}\bra{j}\right)
\end{align*}
and a two-qudit operation $${\sf XX} (\chi) =\exp(-i\chi\sigma^{01}_x\otimes\sigma^{01}_x).$$
The operator $\sigma_\phi^{0j} = \sigma_x^{0j}\cos\phi + \sigma_y^{0j} \sin\phi$ acts within the two-level subspace spanned by $\ket{0}$ and $\ket{j}$, where $\sigma_\kappa^{0j}$ (with $\kappa = x, y$) denotes the Pauli matrix in that subspace. Operations $R_{\phi}^{01}(\theta)$ are performed with laser pulses at 435~nm individually addressing ions. The gate $R_{\phi}^{02}(\theta)$ can be performed only at all ions simultaneously using a microwave antenna at 12.6~GHz. Their fidelity was characterized via qubit-like randomized benchmarking~\cite{knill2008randomized} to be 0.99946(6) and 0.9994(1), respectively. The $R_{z}^{j}(\theta)$ gates are performed virtually~\cite{McKay2017}. Two-qudit ${\sf XX} (\chi)$ gates are implemented with a M$\o$lmer-S$\o$rensen technique~\cite{Molmer-Sorensen1999-2} with a compensation of the single-qudit phase acquired by the spectator $|2\rangle$ state~\cite{nikolaeva2025scalable}. It can be performed on any qudit pair with a mean gate fidelity of 0.963(2) estimated by the measurement of the Bell state preparation fidelity. 

In this work we applied CSB method to two variants of the Toffoli gate decomposition into native gates. One of them used only two states in each ion (qubit implementation), while the second one adopted the approach from~\cite{nikolaeva2025scalable}, where a third state in each particle was used as a clean ancilla reducing required number of two-qudit gates (qutrit implementation). The exact decompositions for both variants are presented in Appendix (Fig.~\ref{fig:Toffoli_schemes}). 

After processing CSB experimental data for both implementations, we obtained sets $\{\lambda_{\tilde{\mathcal{U}}}^{i}\}$ that were converted to diagonal noise matrix elements $\{\lambda_{\mathcal{E}}^i\}$. Distributions of fidelity estimates were subsequently generated from these sets.

\begin{figure}[!htb]
    \centering
    \includegraphics[width=0.98\linewidth]{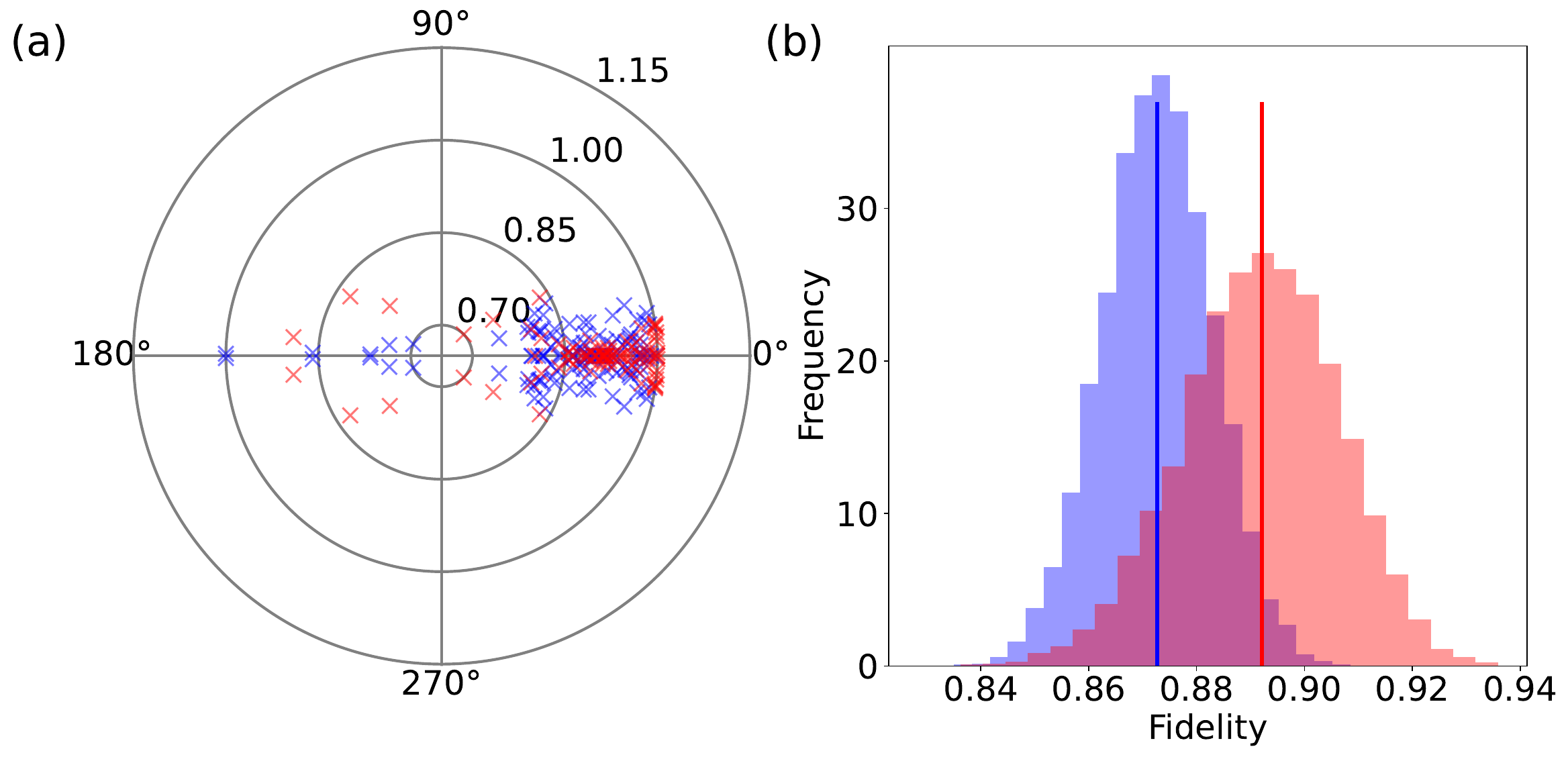}
    \caption{CSB experimental results for qubit (blue) and qutrit (red) Toffoli implementations. (a) Filtered noisy eigenvalues in polar coordinates; (b) Distributions of fidelity estimates from 64 eigenvalues.}
    \label{fig:lambd_boot}
\end{figure}

Fig.~\ref{fig:lambd_boot} displays the obtained $\{\lambda_{\tilde{\mathcal{U}}}^{i}\}$ sets and corresponding fidelity estimate distributions. For the qubit implementation, we obtained $\text{FEI}_{\text{qubit}} = [0.854;0.894]$ (single point fidelity estimate $\hat{\mathcal{F}}_{\text{qubit}} = 0.873$). The qutrit implementation yielded $\text{FEI}_{\text{qutrit}} = [0.866;0.922]$ (single point fidelity estimate $\hat{\mathcal{F}}_{\text{qutrit}} = 0.892$). Notably, the qutrit experiment revealed significant leakage to the qutrit level, which contradicts the Markovian noise assumption. Investigation of such leakage effects represents an important direction for future research.

\section{Conclusion}\label{sec:Conclusion_and_discussion}

This work presents the CSB protocol for estimating fidelity estimate interval (FEI) of quantum operations and its application to the three-qubit Toffoli gate. The protocol demonstrates universality, allowing to analyze operations with degenerate spectrum and remains robust to SPAM errors. Numerical simulation of CSB for the Toffoli gate, conducted with account of amplitude damping noises and unitary perturbations, confirmed effectiveness of the method. Results show that the proposed algorithm produces FEI whose midpoint closely approximates the theoretical fidelity value.

The experimental part of the work included implementation of CSB for two variants of the Toffoli gate on ion platform - qubit and qutrit. According to results of data processing, FEIs and single point fidelity estimates were obtained: $\text{FEI}_{\text{qubit}} = [0.854;0.894]$ (single point fidelity estimate $\hat{\mathcal{F}}_{\text{qubit}} = 0.873$) and $\text{FEI}_{\text{qutrit}} = [0.866;0.922]$ (single point fidelity estimate $\hat{\mathcal{F}}_{\text{qutrit}} = 0.892$).

Despite the advantages, the method has limitations related to dependence of accuracy on the number of experiments $N$ and the range $L_{\max}$. Perspectives of further research include extension of CSB to multi-qubit systems with non-local noises, integration with methods of quantum error correction and adaptation to other platforms such as superconducting qubits.

Thus, the CSB protocol represents an effective tool for diagnostics of quantum operations, combining simplicity of implementation with high accuracy. Its successful application to the Toffoli gate opens possibilities for using the method in a wide class of quantum benchmarking tasks.

\begin{acknowledgments}
The work was supported by Rosatom in the framework of the Roadmap for Quantum computing (Contracts No. 868-1.3-15/15-2021 dated October 5, 2021 and No. 868.1653-D dated August 21, 2025).

The authors are grateful to Alina Mardanova for valuable discussions and comments.
\end{acknowledgments}

\bibliographystyle{apsrev4-2}
\bibliography{bibliography}

\appendix

\section{Quantum circuits for CSB protocol}

\subsection{State preparation circuits}

\begin{figure*}[h]
    \centering
    \includegraphics[width=0.98\linewidth]{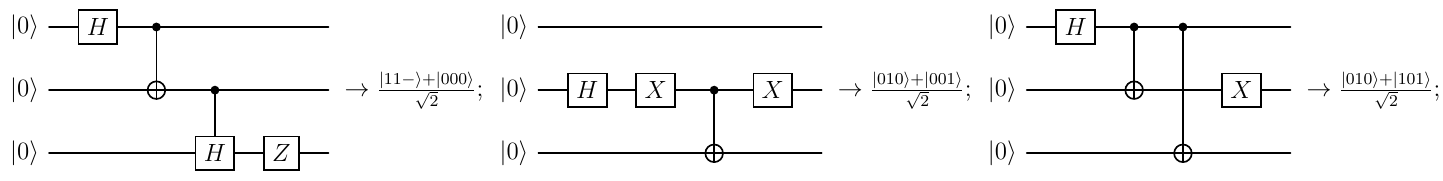}
    \caption{Possible quantum circuits for preparing various non-trivial states for Toffoli gate benchmarking protocol. The circuits can be transpiled using native trapped-ion gate set.}
    \label{fig:state_prep_schemes}
\end{figure*}

Within the CSB protocol, it is necessary to prepare superposition states $|\psi_{ab}\rangle = (|\psi_a\rangle + |\psi_b\rangle)/\sqrt{2}$, where $|\psi_{a}\rangle$ and $|\psi_{b}\rangle$ are eigenstates of gate $U$. In general case, preparing such states is not always efficient. However, for three-qubit Toffoli gate, circuits for preparing such states are known. Fig.~\ref{fig:state_prep_schemes} shows some of them.

\subsection{Toffoli gate decompositions}

\begin{figure*}[h]
    
    \centering
    \includegraphics[width=0.98\linewidth]{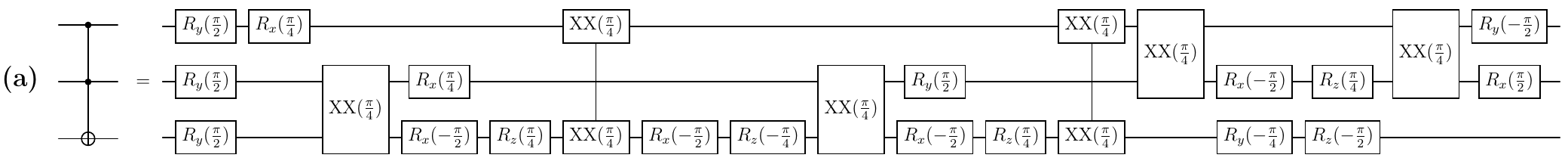}
    
    \vspace{0.5cm}
    
    \includegraphics[width=0.98\linewidth]{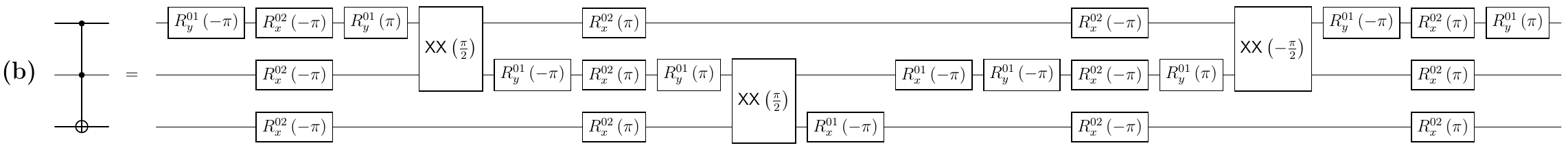}

    \caption{Ion-specific transpilation of the three-qubit Toffoli gate circuit. (a) Qubit-based decomposition, (b) qutrit-based. }
    
    \label{fig:Toffoli_schemes}
\end{figure*}

In the experiment for determining fidelity of the Toffoli gate, 2 different trapped-ion specific decompositions of the Toffoli gate shown in Fig.~\ref{fig:Toffoli_schemes} were compared. The well-known qubit decomposition includes 6 two-qubit $\text{XX}$ operations, while the qutrit one~\cite{nikolaeva2022decomposing,nikolaeva2025scalable} -- only 3. Using the qutrit decomposition allowed to achieve higher fidelity of the Toffoli operation.

\onecolumngrid\
\twocolumngrid\

\onecolumngrid\
\twocolumngrid\

\end{document}